\input amstex 
\input amsppt.sty
\input epsf
\input crossref
\nologo
\pagewidth{160truemm}
\pageheight{247truemm}
\TagsOnRight

\def\ch{\gamma}
\def\ph{\rho}
\def\qh{\sigma}
\def\ab{A}
\def\bb{B}
\def\cb{C}
\let\cal=\Cal
\def\cA{{\cal A}}
\def\cD{{\cal D}}
\def\cE{{\cal E}}
\def\cG{{\cal G}}

\def\cN{{\cal N}}

\def\cR{{\cal R}}

\def\Real{{\Bbb R}}
\def\Nat{{\Bbb N}}
\def\Cplx{{\Bbb C}}
\def\compact{\subset\subset}
\def\normeq{\buildrel * \over =}
\def\pbyp#1#2{{\partial{#1}\over\partial{#2}}}

\def\pbypp#1#2#3{{\partial^2{#1}\over\partial{#2}\partial{#3}}}
\def\sob#1{{\cal H}^{#1}}
\def\covnorm#1#2#3{{\Vert#3\tilde\Vert^{#1}_{#2}}}
\def\sobnorm#1#2#3{{\Vert#3\Vert^{#1}_{#2}}}
\def\intnorm#1#2#3{{\Vert#3\hat\Vert^{#1}_{#2}}}
\def\dn#1{d^{#1}\!}

\def\phin#1{\varphi^{(#1)}} 
\def\EPS#1{{#1}_\eps{}}
\def\EPSUP#1#2{{#1}^{#2}_\eps{}}
\def\EPSDOWN#1#2{{#1}_{#2}^\eps{}}
\def\geps{\EPS{g}}
\def\eueps{\EPSUP{e}}
\def\gdeps{\EPSDOWN{g}}
\def\gueps{\EPSUP{g}}
\def\ueps{\EPS{u}}
\def\veps{\EPS{v}}
\def\weps{\EPS{w}}
\def\feps{\EPS{f}}
\def\boxeps{\square^\eps}
\def\fplus{f_\eps^+{}}
\def\uplus{u_\eps^+{}}
\def\mueps{\mu^\eps}
\def\E#1{{\roman{E}}_\eps^{#1}{}}
\def\T#1#2{T{}^{#2}_{\eps,#1}{}}
\def\green#1#2{G^{#1}_{#2}{}}
\def\greeneps#1#2{G^{#1}_{\eps,#2}{}}
\def\nabeps#1{\nabla^{\eps}_{#1}}
\def\nabepsup#1{\nabla_{\eps}^{#1}}
\def\nabepsdots#1#2#3{\nabeps{#1_{#2}}\ldots\nabeps{#1_{#3}}}
\def\partialdots#1#2#3{\partial_{#1_{#2}}\ldots\partial_{#1_{#3}}}
\def\idxsum#1#2#3{\sum_{#1_{#2}\ldots#1_{#3}}}
\def\idxsumover#1#2#3#4{\sum_{\scriptstyle#1_{#2}\ldots#1_{#3}\atop#4}}
\def\IP#1#2#3#4{#1\langle#3,#4#1\rangle}
\def\ip{\IP\relax\relax}

\let\leq=\leqslant
\let\geq=\geqslant
\let\eps=\varepsilon
\def\constant{{\roman{constant}}}
\def\rmin{r_{\roman{min}}}
\def\LLoc{L_{\roman{Loc}}}

\topmatter
\title
Generalised hyperbolicity in conical space-times
\endtitle
\author
J.A.Vickers and J.P.Wilson
\endauthor
\address
Faculty of Mathematical studies, University of Southampton,
Southampton SO17 1BJ, UK.
\endaddress
\email
jav\@maths.soton.ac.uk, jpw\@maths.soton.ac.uk
\endemail
\abstract
Solutions of the wave equation in a space-time
containing a thin cosmic string are examined in the context of
non-linear generalised functions. Existence and uniqueness of
solutions to the wave equation in the Colombeau algebra $\cG$ is
established for a conical space-time and this solution is shown to be
associated to a distributional solution. A concept of
generalised hyperbolicity, based on test fields, can be defined for such
singular space-times and it is shown that a conical space-time is
$\cG$-hyperbolic. 
\endabstract
\endtopmatter

\document
\crossreferencing

\head
1. Introduction
\endhead

Weak singularities have for some time been used to model physically
plausible scenarios such as thin cosmic strings, impulsive gravitational
waves and shell crossing singularities.  Such singularities typically admit
a locally bounded metric and well behaved curvature scalars but none the
less are still classified as singularities due to a
low differentiable metric resulting from topological defects.

One such example is the conical singularity, used to model thin cosmic
strings, resulting from  writing down Minkowski space-time in cylindrical
coordinates
$$ ds^2=-dt^2+dr^2+r^2d\phi^2+dz^2 $$
and identifying $\phi=0$ with $\phi=2\pi A$ for $A<1$.

The singular behaviour of this space time becomes apparent when one
parallelly transports a frame around a closed curve around the axis
$\Lambda$ and finds that it undergoes a holonomy of the same magnitude as
the deficit angle $2\pi(1-A)$. For non-singular space-times with a $C^{2-}$
regular metric it may be shown that such holonomies are trivial (See Wilson
and Clarke (1996) for a proof); it is therefore the case that there is a
problem with differentiability.
In suitable Cartesian coordinates
$$ x^0=t,\quad
   x^1=r\cos(\phi/A),\quad
   x^2=r\sin(\phi/A),\quad
   x^3=z
$$
the metric may be written as 
$$
\split
   ds^2=-dt^2 +dz^2 &+ \tfrac12(1+A^2)\bigl((dx^1)^2+(dx^2)^2\bigr)\\
   & +\tfrac12(1-A^2) \biggl({(x^1)^2-(x^2)^2\over (x^1)^2+(x^2)^2}\biggr)
   \bigl( (dx^1)^2-(dx^2)^2 \bigl) \\
   & + \tfrac12(1-A^2) 2 \biggl({2x^1x^2\over (x^1)^2+(x^2)^2}\biggr)
   dx^1\,dx^2  
\endsplit
\tag\tagnum\xlabel{cartcone}
$$
which, although it is locally bounded, admits directional dependent
limits as one approaches the axis. 

The low differentiability of the metric at the axis does not prevent the
curvature being calculated as a distribution. One can not directly 
use conventional
distribution theory techniques because this would involve the evaluation of
ill defined products of distributions which are not well defined (Schwartz,
1954). Geroch and Traschen (1987) introduced a class of regular
metrics for which the distributional curvature is guaranteed to be
well defined  and showed that such a space-time must have  
distributional curvature with support on a submanifold of at most
co-dimension one, which is clearly not the case for the cone. 
Instead regularisation techniques may be used such as those employed
by Balasin and Nachbagauer (1993) and especially those using the framework
of Colombeau's non-linear generalised function theory (Colombeau, 1984)
such as Clarke et al., (1996). Using such techniques it may be shown that
the non-zero components 
of the energy-momentum tensor density are
$$ T^0{}_0 (-g)^{1/2} = T^3{}_3 (-g)^{1/2} = -2\pi(1-A)
   \delta^{(2)}(x^1,x^2) $$  
This form of the energy momentum tensor is what justifies the use of this
kind of space-time to model a thin cosmic string. This suggests that the
interpretation of the curvature as a distribution could be one such
criterion for regarding these quasi-regular singularities as physically
plausible.

Another question relating to the physical importance of a singularity 
arises out of the question of to what extent the
singularity disrupts the evolution of Einstein's equations and therefore
global hyperbolicity.  For space-times with a $C^{2-}$ metric (which
guarantees the existence of unique geodesics) this question is answered by
requiring the space-time to be globally hyperbolic (Penrose, 1979).
There are however a number of space-times with a lower differentiability
for which, although they may violate cosmic censorship, there still may be
well-posed initial-value problems for test fields. This led to the proposal
of the concept of generalised hyperbolicity (Clarke, 1998) in which one
examined the extent to which singularities were obstructions to the
evolution of the wave equation. In order to apply this concept to
space-times with shell crossing singularities Clarke replaced the initial
value problem for the wave equation
$$
\square u=f, \qquad u_{|S}=v, \qquad (n^a\partial_a u)_{|S}=w
$$ by a distributional version obtained by multiplying by a test field
$\psi$ and integrating by parts once (rather than twice as is usual)
to give
$$
\int_{M^+} \!\!\! {\partial_a u}\, {\partial_b\psi}\, g^{ab} \,\mu = 
   -\int_{M^+} \!\!\!\psi f \,\mu - \int_S \psi w \,\mu_{S} 
$$ 
He then said that a space-time was $\square$-globally hyperbolic if the
above equation had a unique solution for all $\psi \in \cD(M)$.
Moreover it was shown that this form of generalised hyperbolicity was
satisfied for a class of curve-integrable space-times and in particular it
was demonstrated for the shell-crossing dust space-times.

In this paper we shall consider the question of generalised hyperbolicity
for the conical space-time~\eqref{cartcone}. Unfortunately this space-time
does not admit a locally square integrable connection, so we are unable to
apply the result of Clarke (1998) directly. Instead we shall follow a
different approach: We shall use Colombeau's generalised functions to
overcome the ambiguities which arise when attempting to multiply
distributions which would arise when considering solutions to the wave
equation in a space-time of such low differentiability. This involves
writing down the Cauchy problem for wave equation in the Colombeau algebra
$\cG(M)$, and proving the existence of a unique generalised function
solution. This will be done for a class of singular metrics which includes
the 4-cone~\eqref{cartcone} (Section 4). Having obtained a unique solution
we shall examine to what extent it is possible to interpret it as a
distribution. We say that a space-time is $\cG$-globally hyperbolic if
there exists a unique solution to the wave equation in the Colombeau
algebra $\cG$ and that this solution is associated to a distribution.  We
therefore demonstrate in this paper that a conical space-time is
$\cG$-hyperbolic.

\head
2. Colombeau's generalised functions
\endhead

We first briefly recall the essential details of Colombeau's theory (see e.g.\
Colombeau, 1984 or Oberguggenberger, 1992 for further details).  We denote
the space of smooth functions with compact support $\cD(\Real^n)$, so that
the space of distributions is $\cD'(\Real^n)$ and we define the following
spaces;

\definition{Definition \defnum}
We define the space of smoothing kernels $\cA_q(\Real^n)$ as the space of all
functions $\varphi^{(n)}\in\cD(\Real^n)$ such that
$$
\alignat2
   &\int_{\Real^n} \!\!\varphi^{(n)}(\xi) \,d^n\xi =1 \\ 
   &\int_{\Real^n}\!\! \varphi^{(n)}(\xi)\xi^k \,d^n\xi = 0 
   &&\qquad \forall k \in \Nat^n
 \quad  \hbox{such that}\quad 1 \leq |k| \leq q
\endalignat
$$ 
where we are using a multi-index notation so that 
$\xi^k=(\xi^1)^{k_1}\ldots(\xi^n)^{k_n}$.
\enddefinition

Given a  function $f\in \LLoc^1(\Real^n)$, we may smooth it by
constructing the following convolution;
$$ \tilde{f}(\varphi_\eps,x) = \int_{\Real^n}\!\!
   f(\xi)\phin{n}_\eps(\xi-x)\,\dn{n}\xi $$ 
where 
$$
   \phin{n}_\eps(\xi)=\eps^{-n}\phin{n}(\xi/\eps)
$$
and for more general distributions $T\in\cD'(\Real^n)$ we have
$$ \tilde{T}(\varphi_\eps,x) = \ip{T}{\tau_x\phin{n}_\eps} $$
where
$$
   \tau_x\phin{n}(\xi) =\phin{n}(\xi-x)
$$

We would like to construct a differential algebra of such smoothings;
We define a base space $\cE(\Real^n)$ as the space of all
functions $\tilde f:\cA_0\times\Real^n\to\Cplx$ which are smooth as a
function of $x$.
Certainly smoothings of distributions are elements of this space; however
one desirable property would be for the smoothing of $f\in C^\infty$ to
coincide with its natural embedding in that
$$ \tilde f(\varphi_\eps,x) = \int f(x) \phin{n}_\eps(\xi-x)\dn{n}\xi $$
and
$$ \hat f(\varphi_\eps,x) = f(x) $$
are equivalent. This is achieved by constructing a space of null functions
$\cN(\Real^n)$ such that $\tilde f-\hat f\in\cN(\Real^n)$ and by working
with a quotient space $\cE(\Real^n)/\cN(\Real^n)$.

\definition{Definition \defnum}
The space of null functions $\cN(\Real^n)$ consists of the functions
${\tilde f}\in\cE(\Real^n)$ such that given $K\compact\Real^n$ and any
collection of indices $p_1,\ldots,p_k$ with $k\in\Nat$ there is some
$N\in\Nat$ and an increasing unbounded sequence $(\gamma_q)$ such that for
each $\varphi\in\cA_q$ for $q\geq N$, $\exists c,\eta>0$ such that
$$ \sup_{x\in K} | \partialdots{p}{1}{k}{\tilde f}(\varphi_\eps,x) |
   \leq c \eps^{\gamma_q-N} \qquad  (0<\eps<\eta) $$ 
\enddefinition

The quotient space $\cE(\Real^n)/\cN(\Real^n)$ is not well defined because
$\cN(\Real^n)$ is not an ideal of $\cE(\Real^n)$. We may multiply a null
function with a function of non-polynomial growth in $\eps^{-1}$ to give a
function which is not null. This problem is rectified by restricting
$\cE(\Real^n)$ to a space of moderate functions $\cE_M(\Real^n)$ which
itself is an algebra and of which $\cN(\Real^n)$ is an ideal. It may be
verified that the smoothings of distributions are indeed moderate.

\definition{Definition \defnum}
The space of moderate functions $\cE_M(\Real^n)$ consist of the functions
$\tilde f\in\cE(\Real^n)$ such that given $K\compact\Real^n$ and any
collection of indices $p_1,\ldots,p_k$ with $k\in\Nat$ there is some
$N\in\Nat$ such that for each $\varphi\in\cA_N$, $\exists c,\eta>0$ such that
$$ \sup_{x\in K} | \partialdots{p}{1}{k}{\tilde f} (\varphi_\eps,x) |
   \leq c \eps^{-N} \qquad (0<\eps<\eta) $$ 
\enddefinition

We finally define our space of generalised functions as the quotient space
$$ \cG(\Real^n) = { \cE_M(\Real^n) \over \cN(\Real^n)} $$
The space $\cG(\Real^n)$ is a differential algebra, of which
$C^\infty(\Real^n)$ is a subalgebra and $\cD'(\Real^n)$ is a linear
subspace. In practice one usually works with the representative moderate
functions to perform calculations in this space.  Many elements of
$\cG(\Real^n)$ do not result as the smoothing of distributions, but even so
are equivalent to a distribution in the following sense;

\definition{Definition \defnum}
We say that $[\tilde f]\in\cG(\Real^n)$ has an associated distribution
$T\in\cD'(\Real^n)$ (written as $[\tilde f]\approx T$) if for any
representative $\tilde f$ and $\psi\in\cD(\Real^n)$ there exists some
$N\in\Nat$ such that
$$ \lim_{\eps\to0} \int\tilde{f}(\varphi_\eps,x) \psi(x) \,\dn{n}x =
   \ip{T}{\psi} $$
\enddefinition

It may be shown, for example,that if $\tilde H$ and $\tilde\delta$ 
are the smoothed
Heaviside and delta distributions respectively then $[\tilde H]^2\approx H$
and $[\tilde H][\tilde\delta]\approx\tfrac12\delta$. On the other hand, the
generalised function $[\tilde\delta]^2$ has no associated distribution.

When one applies the theory to initial value problems one usually
makes the additional requirement that the $n$-dimensional smoothing
kernels are Cartesian products of one dimensional kernels 
(Oberguggenberger 1992) so that
$$
\phin{n}(x^1,\ldots,x^n)=\phin{1}(x^1)\cdots\phin{1}(x^n).
$$
The reason for doing this is that it gives a natural way of defining
Cartesian products and restrictions. However in order to simplify the
calculations in Section 5 it is useful to work with rotationally
symmetric smoothing kernels. Unfortunately it is not possible to write
such a kernel as a finite sum of Cartesian products and for this
reason we describe in Appendix A an alternative way of restricting
4-dimensional smoothing kernels to an initial surface $S$ so that any
spatial symmetries are preserved. Note however that the results of the
next two sections are not sensitive to the precise restriction process
adopted.

\head
3. The wave equation in the Colombeau algebra $\cG(M)$
\endhead         

We first introduce the convention that lower case Latin indices $a,\ b,\
\ldots $ run over $0 \ldots 3$, Greek indices $\alpha,\ \beta,\
\ldots$ run over $1 \ldots 3$ and  upper case indices $A,\  B,\ \ldots$
run over $1$ and $2$.
Suppose we have a space-time, equipped with a locally bounded singular
metric, $(M,g)$ and we wish to solve the Cauchy problem for the wave
equation 
$$
\alignedat2
  \square u(t,x^\alpha) &= 0  \\
  u(0,x^\alpha) &= v(x^\alpha) \\
  {\partial_t u}(0,x^\alpha) &= w(x^\alpha) \\
  \endalignedat
$$  
where initial data $(v,w)$ lying in the Sobolev spaces  
$\sob1(S)\times \sob0(S)$ is prescribed on the
initial surface $S$, $t=0$.  If a solution exists one would expect it to be
defined as a distribution. This however will cause difficulty in
interpreting
$$ \square u = (-g)^{-1/2} \partial_a \bigl( (-g)^{1/2} g^{ab} \partial_b u
   \bigr) $$   
as a distribution in the framework of Schwartzian distribution theory
because it contains products of $u$ with a singular metric and its weak
derivatives which are not well defined in this theory. 
The way to overcome these shortcomings is to employ
the non-linear generalised function theory of Colombeau~(1984), which does
give a distributional interpretation to many distributional products that
would otherwise be undefined. We first embed the metric $g_{ab}$ into the
generalised function space $\cG(M)$ by constructing its representative
$(\gdeps{ab})\in\cE_M(M)$ as the convolution integral
$$ \gdeps{ab}(t,x^\alpha)=\int g_{ab}(t+\eps \zeta,x^\alpha+\eps \xi^\alpha)
    \phin{4}(\zeta,\xi^\alpha)
   \,d\zeta\,d{}^3\xi $$ 
Since the initial data $(v,w)$ does not have to be smooth, we must also
embed it into the space $\cG(S)\times\cG(S)$ as $(V,W)$ by defining the
convolution integrals $\veps$ and $\weps$ in $\cE_M(S)$;
$$
\aligned
  \veps(x^\alpha) &= \int v(x^\alpha+\eps \xi^\alpha)
  \phin{3}(\xi^\alpha) \,\dn{3}\xi \\
  \weps(x^\alpha) &= \int w(x^\alpha+\eps \xi^\alpha)
  \phin{3}(\xi^\alpha) \,\dn{3}\xi 
\endaligned
$$
The Generalised function wave operator
$\square:\cG(M)\to\cG(M)$ then may be defined by denoting
$\square U$ for $U\in\cG(M)$ as the class represented by
$(\boxeps \ueps)\in\cE_M(M)$ where
$(\ueps)\in\cE_M(M)$ is a representative for $U$ and
$$ \boxeps \ueps = (-\geps)^{-1/2}
   \partial_a\bigl( (-\geps)^{1/2} \gueps{ab} \partial_b \ueps \bigr) $$
We would like to then be able to solve the  Cauchy problem in the space
$\cG(M)$ 
$$
\alignedat2
  \square U(t,x^\alpha) &= 0  \\
  U(0,x^\alpha) &= V(x^\alpha) \\
  {\partial_t U}(0,x^\alpha) &= W(x^\alpha) \\ 
\endalignedat
\tag\tagnum\xlabel{cauchy1}
$$
and obtain a solution $U\in\cG(M)$ which is associated to a distribution.
In practice one works with the equivalent problem in $\cE_M(M)$;
$$
\alignedat2
  \boxeps \ueps(t,x^\alpha)  &= \feps(t,x^\alpha)    \\
  \ueps(0,x^\alpha) &= \veps(x^\alpha)  \\
  {\partial_t\ueps}(0,x^\alpha) &= \weps(x^\alpha)  \\
\endalignedat
\tag\tagnum
\xlabel{cauchy2}
$$
where $(\feps)\in\cN(M)$ and obtain the associated distribution (if
there is one) $u$ by evaluating the limit for each $\psi\in\cD(M)$
$$ \ip{u}{\psi}=\lim_{\eps\to0} \int \ueps\bigl(\varphi,(t,x^\alpha)\bigr)
   \psi(t,x^\alpha) \mueps(t,x^\alpha) \tag\tagnum\xlabel{assoc.limit}$$ 
where $\varphi\in\cA_q$ for $q\in\Nat$ large enough.

The question of existence and uniqueness of solutions to~\eqref{cauchy1} is
not as straight-forward as for its classical function
counterpart. Although  one may
establish the existence of unique solutions for~\eqref{cauchy2} (for each
$\varphi\in\cA_0$ and $\eps>0$) by using the using the
Cauchy-Kowalewska theorem for the case of analytic data, this is not
enough to establish the existence of unique solutions
to~\eqref{cauchy1}. One must carefully examine the role that moderate and
null functions have to play. (See Oberguggenberger~(1989) for examples of
similar Cauchy problems that do not admit unique solutions). Firstly, we
want $U$ to be an element of $\cG(M)$. This means that we have to
verify that $(\ueps)\in\cE_M(M)$.  Secondly for $U$ to be a unique we
only require that the solution $(\ueps)$ of~\eqref{cauchy2} are unique up
to an element of $\cN(M)$. This means that for a particular choice of
$(\feps)$, $(\veps)$ and $(\weps)$ we are allowed to have more than one
solution provided they are all well defined elements of $\cE_M(M)$ and
differ by elements of $\cN(M)$.  We must also take into account that
$(\feps)$, $(\veps)$ and $(\weps)$ could have been chosen with freedom up to
a null function, so we require that the solution $(\ueps)$ is augmented by
at most a null function if $(\feps)$, $(\veps)$ and $(\weps)$ are augmented
by null functions.  Since this Cauchy problem is linear it follows that the
solution $U\in\cG(M)$ is unique if for the problem~\eqref{cauchy2},
with $(\veps),(\weps)\in\cN(S)$ implies that
$(\ueps)\in\cN(M)$.

Moreover even if there is a unique solution $U\in\cG(M)$, it need not
admit an associated distribution because the limit~\eqref{assoc.limit} may
be dependent on the smoothing kernel $\varphi\in\cA_q$ or may not even be
defined, however large we may choose $q$.

\head
4. Proof of the existence of unique solutions in $\cG(M)$
\endhead

We start by observing that we may take a representation in $\cE_M(M)$
of the embedded metric which admits the form; 
$$ {ds_\eps}^2=-dt^2+dz^2+\gdeps{\ab\bb}(x^\cb)\,dx^\ab\,dx^\bb
   \tag\tagnum\xlabel{smoothmetric}
$$
where $\gdeps{\ab\bb}\in\cE_M(M)$ and locally
$\gdeps{\ab\bb}$ and its derivatives may be bounded as follows; 
$$
\aligned
   |\gdeps{\ab\bb}|  &\leq M_0 \\
   |\gueps{\ab\bb}| &\leq M_0 \\ 
   |\partial_{\cb_1}\ldots\partial_{\cb_k}\gdeps{\ab\bb}|
   &\leq {M_k\over\eps^{k}} 
\endaligned
$$
with $M_k$ being positive constants independent of $\eps$.  It is
easily seen that on embedding the conical metric~\eqref{cartcone} into
$\cG(M)$, the $\gdeps{\ab\bb}$ and its derivatives admit this form;
$$
\aligned
 \gdeps{11}(x^\ab) &= \tfrac12(1+A^2) + \tfrac12(1-A^2) P_\eps(x^\ab) \\
 \gdeps{12}(x^\ab) &= \tfrac12(1-A^2) Q_\eps(x^\ab) \\
 \gdeps{22}(x^\ab) &= \tfrac12(1+A^2) - \tfrac12(1-A^2) P_\eps(x^\ab)
\endaligned
$$
where
$$
\aligned
  P_\eps(x^\ab) &= {1\over\eps^2} \int {(\xi^1)^2-(\xi^2)^2\over
  (\xi^1)^2+(\xi^2)^2 }\varphi^{(2)}
  \biggl({\xi^{\ab}-x{}^{\ab}\over\eps}\biggr) \,\dn{2}\xi \\
  Q_\eps(x^\ab) &= {1\over\eps^2} \int 2{\xi^1\xi^2\over
  (\xi^1)^2+(\xi^2)^2} 
  \varphi^{(2)}\biggl({\xi^{\ab}-x{}^{\ab}\over\eps}\biggr)\,\dn{2}\xi 
\endaligned
$$
and $\varphi^{(2)}$ is a two dimensional smoothing kernel obtained by
integrating out the $t$ and $z$ dependence.

Our aim is to estimate solutions $\ueps$ of~\eqref{cauchy2} and its
derivatives in terms of powers of $\eps$, given the moderate and null
bounds of $\feps$, $\veps$ and $\weps$ using a method of energy estimates
following Hawking and Ellis~(1973) and Clarke~(1998).  We shall assume that
$K$ is a compact region of $M$ which intersects the initial hypersurface
$S=S_0$ described by $t=0$ and that and $\{S_\tau\}_{0\leq\tau\leq\tau_1}$
is a family of smooth space-like hypersurfaces in $K$ which intersect $S_0$
on a common 2-surface. See Figure 1.

\midinsert
\medskip
\centerline{\epsfbox{pic1.ps }}
\medskip
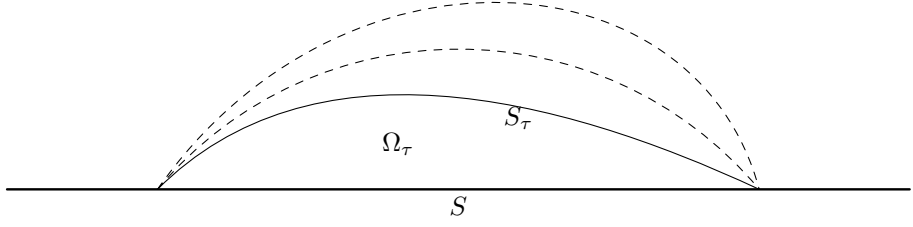
\centerline{\it Figure 1.}
\medskip
\endinsert

We shall assume that the surfaces $S_\tau$ may be
expressed by the equation $\sigma(t,x^\alpha)=\tau$; we may therefore define a
normal to $S_\tau$, $n_a=\partial_a\tau$ and a surface element
$\mueps_{S_\tau}$ which satisfies
$$ \mu_\eps = (-\geps)^{1/2} \,d^4x = \mueps_{S_\tau} \wedge d\tau $$ 
We shall write the region bounded by $S_0$ and $S_\tau$ as
$$ \Omega_\tau = \bigcup_{\zeta\in[0,\tau]} S_{\zeta} $$

Since we will have to estimate $\ueps$ and its derivatives of arbitrary
order, we shall construct a hierarchy of energies. It will turn out that it
is much more convenient to work with energy integrals based on covariant
derivatives rather than classical Sobolev norms based on partial
differentiation.  We define our energy integrals by the formula
$$ \E{k}(\tau)=\sum^k_{j=0} \int_{S_\tau} \T{j}{ab}\, t_b \,
    n_a \,\mueps_{S_\tau} \tag\tagnum\xlabel{energies}$$
where
$$
\aligned
   \T{k}{ab} &=
   \bigl(\gueps{ac} \gueps{bd} -\tfrac12 \gueps{ab} \gueps{cd}\bigr)
   \eueps{p_1q_1}\ldots\eueps{p_{k-1}q_{k-1}} \,
   \nabeps{c}\nabepsdots{p}{1}{k-1}{\ueps}\,
   \nabeps{d}\nabepsdots{q}{1}{k-1}{\ueps} \\
   \T{0}{ab} &= -\tfrac12 \gueps{ab} {\ueps}^2 \\
   t_a &= \partial_a t \\
   \eueps{ab} &= \gueps{ab} + 2 t^a t^b
\endaligned
$$
The form of the metric~\eqref{smoothmetric} guarantees that
$\nabeps{a}t_b=0$ This enables the tensor $\eueps{ab}$ to be constructed so
that it is positive definite and is annihilated by covariant
differentiation i.e. $\nabeps{a}\eueps{bc}=0$.

We also define the following 3-dimensional Sobolev norms
$$
\aligned
   \sobnorm{k}{S_\tau}{\ueps} &=
   \biggl\{ \idxsumover{p}{1}{j}{0\leq j \leq k}
   \!\int_{S_\tau}\! |\partialdots{p}{1}{j}\ueps|^2
   \,\mueps_{S_\tau} \biggr\}^{1/2} \\
   \covnorm{k}{S_\tau}{\ueps} &=
   \biggl\{ \idxsumover{p}{1}{j}{0\leq j \leq k}
   \!\int_{S_\tau}\! |\nabepsdots{p}{1}{j}\ueps|^2
   \,\mueps_{S_\tau} \biggr\}^{1/2}
\endaligned
$$
These are equivalent to the energy integrals in the following sense

\proclaim{Lemma \procnum}
\xlabel{sandwich}
There exist positive constants $A_k$, $A'_k$, $B_k$ and $B'_k$ such that
$$ \alignat1
   \E{k}(\tau) &\leq A_k \bigl( \covnorm{k}{S_\tau}{\ueps} \bigr)^2
   \tag\tagnum\xlabel{sandwich1}\\
   \bigl( \covnorm{k}{S_\tau}{\ueps} \bigr)^2 &\leq A'_k \E{k}(\tau)
   \tag\tagnum\xlabel{sandwich2}\\
   \bigl( \covnorm{k}{S_\tau}{\ueps} \bigr)^2 &\leq B'_k \sum_{j=1}^k
   {1\over\eps^{2(k-j)}} \bigl( \sobnorm{j}{S_\tau}{\ueps} \bigr)^2 
   \tag\tagnum\xlabel{sandwich3}\\
   \bigl( \sobnorm{k}{S_\tau}{\ueps} \bigr)^2 &\leq B_k \sum_{j=1}^k
   {1\over\eps^{2(k-j)}} \bigl( \covnorm{j}{S_\tau}{\ueps} \bigr)^2 
   \tag\tagnum\xlabel{sandwich4}
\endalignat
$$
\endproclaim

$\sobnorm{k}{S_\tau}{\ueps}$, $\covnorm{k}{S_\tau}{\ueps}$ and $\E{k}(\tau)$
are functions of $\tau$ and $\varphi$. We now define the concepts of having
a moderate or null bound for such an object.

\definition{Definition \defnum}
We say that $\rho:\cA_0\to\Cplx$ has a moderate bound if  there is some
$N\in\Nat$ such that for each $\varphi\in\cA_N$, $\exists c,\eta>0$ such that
$$ | \rho(\varphi_\eps) | \leq c \eps^{-N} \qquad (0<\eps<\eta) $$ 
\enddefinition

\definition{Definition \defnum}
We say that $\rho:\cA_0\to\Cplx$ has a null bound if there is some
$N\in\Nat$ and an increasing unbounded sequence $(\gamma_q)$ such that for
each $\varphi\in\cA_q$ for $q\geq N$, $\exists c,\eta>0$ such that
$$ | \rho(\varphi_\eps) | \leq c \eps^{\gamma_q-N} \qquad  (0<\eps<\eta) $$ 
\enddefinition

An important consequence of Lemma~\xref{sandwich} is that $\E{k}(\tau)$ has
a moderate bound if and only if $\sobnorm{k}{S_\tau}{\ueps}$ has a moderate
bound and that $\E{k}(\tau)$ has a null bound if and only if
$\sobnorm{k}{S_\tau}{\ueps}$ has a null bound. It is therefore sufficient
to estimate these energy integrals.  The idea is to form an energy
inequality to give a bound on $\E{k}(\tau)$ in terms of the lower order
energies $\E{1}(\tau),\ldots,\E{k-1}(\tau)$ and $\E{1}(0),\ldots,\E{k}(0)$,
which is determined by the initial data, and to show that the properties of
the $\E{k}(0)$ having a moderate (null) bound may be carried through to
$\E{k}(\tau)$.

We first integrate around the region $\Omega_\tau$ and apply Stokes'
theorem to obtain
$$ \E{k}(\tau)-\E{k}(0)=\sum^k_{j=0} \int_{\Omega_\tau}
    \!t_b\, \nabeps{a}\T{k}{ab} \,\mueps  $$  
where
$$
\split
  \nabeps{a}\T{k}{ab} &= \eueps{p_1q_1}\ldots\eueps{p_{k-1}q_{k-1}} \,
  \gueps{ac} \gueps{bd} \,
  \nabeps{a}\nabeps{c}\nabepsdots{p}{1}{k-1}\ueps\,
  \nabeps{d}\nabepsdots{q}{1}{k-1}\ueps \\
  & \phantom{={}} +2 \eueps{p_1q_1}\ldots\eueps{p_{k-1}q_{k-1}} \,
  \gueps{ab} \gueps{cd} \,
  \nabeps{c}\nabepsdots{p}{1}{k-1}\!\ueps
  \,\nabeps{[a}\nabeps{d]}\nabepsdots{q}{1}{k-1}\!\ueps \\
  \nabeps{a}\T{0}{ab} &= - \nabepsup{b}\ueps\, \ueps
\endsplit
\tag\tagnum
\xlabel{divT}
$$
This expression involves derivatives of order $k+1$, but we may eliminate
these substituting the wave equation into it. For the base energy case
$k=1$;
$$
\aligned
  \nabeps{a}\T{1}{ab} &=  \feps \, \nabepsup{b}\ueps \\
  \nabeps{a}\T{0}{ab} &= -\ueps \, \nabepsup{b}\ueps  
\endaligned
$$
Therefore

$$ \E{1}(\tau) = \E{1}(0) + \int_{\Omega_\tau} \! t^a \,
   \nabeps{a}\ueps\,(\feps-\ueps)\,\mueps $$ 
On making estimates;
$$
\split
  \E{1}(\tau) &\leq \E{1}(0) + L_1 \covnorm{1}{\Omega_\tau}{\ueps}
  \bigl( \covnorm{0}{\Omega_\tau}{\feps} 
  + \covnorm{0}{\Omega_\tau}{\ueps} \bigr) \\  
  & \leq \E{1}(0) + \tfrac12 L_1
  \bigl( \covnorm{0}{\Omega_\tau}{\feps} \bigr)^2 
  + \tfrac32 L_1 \bigl( \covnorm{1}{\Omega_\tau}{\ueps} \bigr)^2
\endsplit
$$
where
$$
\split
   \covnorm{k}{\Omega}{\ueps}
   &= \biggl\{ \idxsumover{p}{1}{j}{0\leq j\leq k}
    \int_{\Omega}|\nabepsdots{p}{1}{j}\ueps|^2 \,\mueps
   \biggr\}^{1/2} \\
   &= \biggl\{\int_{\zeta=0}^\tau \bigl( \covnorm{k}{S_{\zeta}}{\ueps}
   \bigr)^2 \,d\zeta \biggr\}^{1/2}
\endsplit
$$
On applying~\eqref{sandwich2} this inequality  may be written as
$$ \E{1}(\tau) \leq \E{1}(0) + \tfrac12 L_1 \bigl(
   \covnorm{0}{\Omega_\tau}{\feps} \bigr)^2  + \tfrac32 L_1 A'_1
   \int_{\zeta=0}^\tau \!\!\! \E{1}(\zeta) \,d\zeta 
   \tag\tagnum\xlabel{Eineq1} $$

An energy inequality may be also obtained for higher order energies; we
must first eliminate the derivatives of order $k+1$ from the
expression~\eqref{divT} by using the differentiated wave equation
$$ \gueps{ab} \nabepsdots{p}{1}{k-1}\nabeps{a}\nabeps{b}\ueps
   = \nabepsdots{p}{1}{k-1}\feps $$ 
In order to substitute this into~\eqref{divT} we must shuffle the covariant
derivative indices of $u$ by repeatedly applying the Ricci identities
$$ 2\nabla_{[a}\nabla_{b]} X_{p_1p_2\ldots p_k} = R_{abp_1}{}^c\,
   X_{cp_2\ldots p_k} + R_{abp_2}{}^c\, X_{p_1c\ldots p_k} + \cdots +
   R_{abp_k}{}^c\, X_{p_1p_2\ldots c} $$
In this way we find that that
$$
\alignat1
  \gueps{ab}\, \nabeps{a}\nabeps{b}\nabepsdots{p}{1}{k-1}\ueps
   &= \nabepsdots{p}{1}{k-1}\feps  +\sum_{j=1}^{k-1}
  {\cR}^{(k-1,j)}_{p_1\ldots p_{k-1}}\ueps 
  \tag\tagnum\xlabel{shuffle1}\\ 
  2\nabeps{[a}\nabeps{b]}\nabepsdots{p}{1}{k-1}\ueps
  &= \cR^{(k-1,k-1)}_{abp_1\ldots p_{k-1}}\ueps 
  \tag\tagnum\xlabel{shuffle2}\\ 
\endalignat
$$
where $\cR^{(k,j)}\ueps$ represents a linear
combination of contractions of the $(k-j)$th covariant derivative of the
Riemann tensor with the $j$th covariant derivative of $\ueps$. This
quantity may be bounded as
$$ |\cR^{(k,j)}\ueps|^2 \leq {C_{k,j}\over\eps^{2(2+k-j)}}
   \sum_{q_1\ldots q_j} |\nabepsdots{q}{1}{j}\ueps|^2 
$$
where $C_{k,j}$ is a uniform constant.
Now the expressions~\eqref{shuffle1} and~\eqref{shuffle2} may be bounded as
follows;  
$$
\aligned
  |\gueps{ab} \nabeps{a}\nabeps{b}\nabepsdots{p}{1}{k-1}\ueps |^2 
  &\leq k | \nabepsdots{p}{1}{k-1}\feps|^2
  + k \sum_{j=1}^{k-1} |\cR^{(k-1,j)}_{p_1\ldots p_{k-1}}\ueps|^2
  \\\vspace\smallskipamount 
  &\leq C_k  \idxsum{q}{1}{k} |\nabepsdots{q}{1}{k-1}\feps|^2 \\
  &\phantom{{}\leq{}} + C_k \!\!\!\idxsumover{q}{1}{j}{1\leq j\leq k-1}
  {1\over\eps^{2(1+k-j)}} |\nabepsdots{q}{1}{j}\ueps|^2
  \\\vspace\medskipamount 
  |2\nabeps{[a}\nabeps{b]}\nabepsdots{q}{1}{k-1}\ueps|^2
  &\leq {C_k \over \eps^4} \idxsum{q}{1}{k-1}
  |\nabepsdots{q}{1}{k-1}\ueps|^2 
\endaligned
$$
where $C_k$ is a uniform constant.
Therefore
$$
\split
  |\nabeps{a}\T{k}{ab}| &\leq C'_k \biggl\{
  \idxsum{p}{1}{k}\!|\nabepsdots{p}{1}{k}\ueps|^2\biggr\}^{1/2} \\
  &\phantom{\leq} \qquad\qquad \times \biggl\{
  \idxsum{p}{1}{k-1}\! |\nabepsdots{p}{1}{k-1}\feps|^2
  + \!\idxsumover{p}{1}{j}{1\leq j\leq k-1}\!
  {1\over\eps^{2(1+k-j)}}
  |\nabepsdots{p}{1}{j}\ueps|^2 \biggr\}^{1/2} \\
  &\leq \tfrac12 C'_k \biggl\{
  \idxsum{p}{1}{k}\! |\nabepsdots{p}{1}{k}\ueps|^2
  + \!\!\!\idxsum{p}{1}{k-1}\!\!\! |\nabepsdots{p}{1}{k-1}\feps|^2
  + \!\!\!\idxsumover{p}{1}{j}{1\leq j\leq k-1}\!\!\!
  {1\over\eps^{2(1+k-j)}}|\nabepsdots{p}{1}{j}\ueps|^2 \biggr\}
\endsplit
$$
On integrating this becomes
$$ \E{k}(\tau) \leq \E{k}(0) + C''_k \biggl\{
   \bigl(\covnorm{k}{\Omega_\tau}{\ueps}\bigr)^2 +
   \bigl(\covnorm{k-1}{\Omega_\tau}{\feps}\bigr)^2  
   + \sum_{j=1}^{k-1} {1\over\eps^{2(1+k-j)}}
   \bigl(\covnorm{j}{\Omega_\tau}{\ueps}\bigr)^2 \biggr\}
$$
which may be turned into an energy inequality by applying Lemma~\xref{sandwich}
$$
\split
  \E{k}(\tau) \leq \E{k}(0) &+ C'_k
  \bigl(\covnorm{k-1}{\Omega_\tau}{\feps}\bigr)^2  +
   C''_k \int^\tau_{\zeta=0} \!\!\! \E{k} (\zeta)\,d\zeta \\ 
  &+ C'''_k \sum_{j=1}^{k-1} {1\over\eps^{2(2+k-j)}}
  \int^\tau_{\zeta=0} \!\!\! \E{j}(\zeta)\,d\zeta.
\endsplit
\tag\tagnum
\xlabel{Eineq2}
$$

Having obtained this energy inequality we shall prove the following

\proclaim{Lemma \procnum}
\xlabel{solveenergy}
Suppose that $\covnorm{0}{\Omega_\tau}{\feps}, \ldots,
\covnorm{k-1}{\Omega_\tau}{\feps}$ have a null bounds. Then
\roster
  \item  if\/ $\E{0}(0),\ldots,\E{k-1}(0)$ have moderate bounds then so
  does $\E{k}(\tau)$ 
  \item  if\/ $\E{0}(0),\ldots,\E{k-1}(0)$ have null bounds then so does
  $\E{k}(\tau)$ 
\endroster
\endproclaim

\demo{Proof}
For part 1, we proceed by induction (Proof of part 2 is similar). For the
case $k=1$ we may apply Gronwall's Lemma to the energy
inequality~\eqref{Eineq1}
$$ \E{1}(\tau) \leq \Bigl( \E{1}(0) + \tfrac12 L_1 \bigl(
   \covnorm{0}{\Omega_\tau}{\feps} \bigr)^2 \Bigr) e^{3L_1\tau_1/2} $$ 
therefore if for $\varphi\in\cA_q$ ($q\geq N$) it is the case that
$\E{1}(0)=O(\eps^{-N})$ and
$\covnorm{0}{\Omega_\tau}{\feps}=O(\eps^{\gamma_q-N})$ it will also be
the case that $\E{1}(\tau)=O(\eps^{-N})$

Now suppose we have for $\varphi\in\cA_q$ ($q\geq N$);
$$
\xalignat2
  \covnorm{j}{\Omega_\tau}{\feps} &= O(\eps^{\gamma_q-N})
    && j=0\ldots  k-1\\ 
  \E{j}(\tau) &= O(\eps^{-N}) \\
  \E{k}(0)    &= O(\eps^{-N}) \\
\endxalignat
$$
on applying Gronwall's inequality to~\eqref{Eineq2} gives
$$ \E{k}(\tau) \leq \biggl\{ \E{k}(0) + C'_k
   \bigl(\covnorm{k-1}{\Omega_\tau}{\feps}\bigr)^2 + C'''_k \sum_{j=1}^k
   {1\over\eps^{2(2+k-j)}} \int^\tau_{\zeta=0} \!\!\! \E{j}(\zeta)
   \,d\zeta \biggr\} e^{C'''\tau}
$$  
therefore
$$\E{k}(\tau)=O(\eps^{-(N+2k)})$$
\enddemo

We now want to apply Lemma~\xref{solveenergy} to show that if
$\veps,\,\weps\in\cE_M(S)$ then $\ueps\in\cE_M(M)$ and similarly if
$\veps,\,\weps\in\cN(S)$ then $\ueps\in\cN(M)$.

The initial data $(\veps,\weps)$ may be used to determine the initial
energies $\E{k}(0)$ by differentiating the Cauchy problem~\eqref{cauchy2}
and using the wave equation as follows
$$ \alignat1
   \partialdots{\ph}{1}{k}\ueps(0,x^\alpha)
   &= \partialdots{\ph}{1}{k}\veps(x^\alpha) \\
   \partial_t \partialdots{\ph}{1}{k}\ueps(0,x^\alpha)
   &= \partialdots{\ph}{1}{k}\weps(x^\alpha) \\
   \partial^j_t\partialdots{\ph}{1}{k}\ueps(0,x^\alpha)
   &= \partialdots{\ph}{1}{k}\bigl\{
   \gueps{\alpha\beta} \partial_\alpha\partial_\beta\partial^j_t\ueps + 
   \gueps{\alpha\beta} \Gamma^\ch_{\alpha\beta}
   \partial_\ch\partial^j_t\ueps \bigr\}(0,x^\alpha) \\
   & \qquad \qquad \qquad -
   \partial^{j-2}_t\partialdots{\ph}{1}{k}\feps(0,x^\alpha)   
   \tag\tagnum\xlabel{diffcauchy.3}
\endalignat$$
and then using~\eqref{energies} to define the energy integrals

Equation~\eqref{diffcauchy.3} may be expressed as
$$ \aligned
   \partial^j_t\partialdots{\ph}{1}{k}\ueps(0,x^\alpha) &= \sum_{l=1}^{k+2}
   {G_l}{}^{\qh_1\ldots \qh_l}_{\ph_1\ldots \ph_k}{}(0,x^\alpha)
   \partialdots{\qh}{1}{l}\partial{}^{j-2}_t\ueps(0,x^\alpha) \\
   & \qquad- \partial^{j-2}_t\partialdots{\ph}{1}{k}\feps(0,x^\alpha)
\endaligned$$
where ${G_l}{}^{\qh_1\ldots \qh_l}_{\ph_1\ldots \ph_k}$ is
constructed by sums and products of the metric and its derivatives, hence
is an element of $\cE_M(S)$. It therefore follows (by an inductive
argument) that
\roster
\item $\veps,\,\weps\in\cE_M(S)$ and $\feps\in\cN(M)$ implies that
$\partial^j_t\partialdots{p}{1}{k}\ueps(0,x^\alpha)\in\cE_M(S)$  
\item $\veps,\,\weps\in\cN(S)$ and $\feps\in\cN(M)$ implies that
$\partial^j_t\partialdots{p}{1}{k}\ueps(0,x^\alpha)\in\cN(S)$  
\endroster
We therefore have proved the following
\proclaim{Lemma \procnum}
\xlabel{initialbounds}
Suppose that $(\veps),\,(\weps)\in\cE_M(S)$ and $(\feps)\in\cN(S)$ then
$\E{k}(0)$ has a moderate bound, and if in addition that
$(\veps),\,(\weps)\in\cN(S)$ then $\E{k}(0)$ has a null bound.
\endproclaim
 
We also want to translate the bounds for $\E{k}(\tau)$, as given by
Lemma~\xref{solveenergy} back to bounds for $\ueps$ and its
derivatives. This may be done by applying Lemma~\xref{sandwich} in
conjunction with the Sobolev embedding theorem (Hawking and Ellis, 1973).
Suppose $M$ is a manifold admitting an $n$ dimensional embedded submanifold
$M'$ then according to the theorem given $k>0$, $\exists M_k>0$ such that
$\forall u\in \sob{k+m}(M')$ with $(2m>n)$
$$ |\partialdots{\ph}{1}{k}u| \leq M_k \intnorm{k+m}{M'}{u} 
   \tag\tagnum\xlabel{sobolev}
$$
where
$$ \intnorm{k}{M'}{u} = \biggl\{
   \idxsumover{\ph}{1}{j}{0\leq j \leq k}\!
   \int_{M'}\!|\partialdots{\ph}{1}{j}u|^2
   \,\mu_{S} \biggr\}^{1/2}
   \tag\tagnum\xlabel{sobembed}
$$
and $\sob{k}(M')$ is the Sobolev space of functions $u$ existing almost
everywhere and for which $\intnorm{k}{M'}{u}<\infty$.

It should be carefully observed that this theorem only gives bounds on the
derivatives in the tangential directions to $M'$ and likewise it shows that
they may be bounded above by a Sobolev norm that only involves those
derivatives.

We first of all apply~\eqref{sobolev} to $\ueps$ on the
submanifold $S_\tau$. This gives us for $\ueps$
$$ |\partial_{\ph_1}\ldots\partial_{\ph_{k}}\ueps| 
\leq P_{0,k} \intnorm{k+m}{S_\tau}{\ueps} 
\leq P_{0,k} \sobnorm{k+m}{S_\tau}{\ueps} 
\qquad (2m>3) $$ 
Where the second inequality comes from the fact that  we may replace the
tangential Sobolev norm with the more crude version
$\sobnorm{k}{M'}{u}$ as its upper bound.
However the above result  does not put any bounds on any of 
the time derivatives. We do
this by applying~\eqref{sobolev} to $\partial^j_t \ueps$;
$$
\aligned
  | \partialdots{\ph}{1}{k} \partial^j_t\ueps| &\leq
  P_{j,k} \sobnorm{k+m}{S_\tau}{\partial^j_t\ueps} \\
  &\leq P_{j,k} \sobnorm{j+k+m}{S_\tau}{\ueps}
\endaligned
$$
Therefore we have shown that $\exists M_k>0$ such that
$$ |\partialdots{p}{1}{k}\ueps| \leq M_k
   \sobnorm{k+m}{S_\tau}{\ueps}, \qquad (2m>3) 
   \tag\tagnum\xlabel{sobbound} $$
On combining \eqref{sobbound} with Lemma~\xref{sandwich} we have proved
the following

\proclaim{Lemma \procnum}
\xlabel{finalbounds}
\roster
\item If $\E{k}(\tau)$ has a moderate bound then $\ueps\in\cE_M(M)$
\item If $\E{k}(\tau)$ has a null bound then $\ueps\in\cN(M)$
\endroster
\endproclaim

On combining Lemmas~\xref{solveenergy}, \xref{initialbounds}
and~ \xref{finalbounds} we have proved our main theorem.

\proclaim{Theorem \procnum}
\roster
\item If $\veps,\,\weps\in\cE_M(S)$ and $\feps\in\cN(M)$ then
$\ueps\in\cE_M(M)$ 
\item If $\veps,\,\weps\in\cN(S)$ and $\feps\in\cN(M)$ then
$\ueps\in\cN(M)$ 
\endroster
\endproclaim

We therefore conclude that for metrics of the form~\eqref{smoothmetric}, a
unique solution $U\in\cG(M)$ exists to the Cauchy
problem~\eqref{cauchy1}.

\head
5. The distributional interpretation of the solution
\endhead

It has been established that the solution $\ueps$ may be interpreted as a
representative for an element of $\cG(M)$. We would like to now discuss
whether or not it may be interpreted as a distribution. Thus we want
$\ueps$ to be associated to a distribution $u$; that is for each
$\psi\in\cD(M)$, $\exists q\in\Nat$ such that the limit
$$ \ip{u}{\psi} = \lim_{\eps\to0} \int_M\! \ueps(\varphi,x) \psi(x)
   \mueps(x) \tag\tagnum\xlabel{assoc} $$
is well defined and independent of $\varphi$.

To do this we consider a representation for $\ueps$ in terms of a
Greens function. Assuming that $\ueps=0$ for $t<0$, we may write the 
Cauchy problem in an
`inhomogeneous' form
$$ \boxeps \uplus = \fplus $$
where
$$
\aligned
  \uplus(t,x^\alpha) &= H(t) \ueps(x^\alpha) \\
  \fplus(t,x^\alpha) &= H(t) \feps(x^\alpha) - \delta'(t) \veps(x^\alpha)
     - \delta(t) \weps(x^\alpha)
\endaligned
$$
In this way one may express the solution in terms of a Greens function 
(see Appendix B)
$$ \uplus(x) = \int \greeneps+{x}(\xi) \fplus(\xi) \mueps $$
It is therefore the case that if the limit~\eqref{assoc} exists it
will be given by
$$ \ip{u}{\psi} = \lim_{\eps\to0} \int_M \! \uplus(\varphi,x) \psi(x)\mueps  
   \tag\tagnum\xlabel{lim}
$$

On expressing $\uplus$ in terms of the Greens function and using the self
adjoint property of $\square$ together with the fact that $G^+$
is integrable on $M \times M$, the order of integration may be
interchanged giving
$$
\alignat1
  \int_M \uplus(\varphi,x) \psi(x) \mu
  & = \int_M \biggl( \int_M \greeneps+{x}(\xi) \fplus(\xi) \mu(\xi) \biggr)
   \psi(x) \mueps(x) \\
  &= \int_{M \times M} \!\!\! G_\eps^+(x,\xi) \fplus(\xi) \psi(x)
  \mueps(x,\xi) \\ 
  &= \int_{M \times M}\!\!\! G_\eps^-(x,\xi) \fplus(x) \psi(\xi)
  \mueps(x,\xi) \\ 
  &= \int_M \lambda_\eps(x) \feps(x) \mueps(x) \\
  & \qquad - \int_S \bigl( \partial_t \lambda_\eps(0,x^\alpha)
   \veps(x^\alpha) + \lambda_\eps(0,x^\alpha) \weps(x^\alpha)
   \bigr) \mueps_S(x^\alpha) \tag\tagnum\xlabel{intupsi}
\endalignat
$$
where
$$ \lambda_\eps(x)= \int_M \greeneps-{x}(\xi) \psi(\xi) \mueps(\xi) $$
Since $\feps$ is null the first term on the right hand side
of~\eqref{intupsi} will vanish as $\eps\to0$ for large enough $q\in\Nat$,
provided that $\lambda_\eps$ admits at most a moderate growth.  For the
second term to admit a well defined limit, independent of
$\varphi\in\cA_q$, as $\eps\to0$, we require that the limiting function of
$\lambda_\eps$ is also well defined and independent of $\varphi$. This is
because $\veps$ and $\weps$ are the embeddings of locally square integrable
functions on $S$.  Thus to show that $\ueps$ is associated to a
distribution we need to show that $\lambda_\eps$ is sufficiently well
behaved.  To do this we follow the approach of Bruhat (1962) and use the
fact that for $\eps >0$, $\lambda_\eps(x)$ is a solution of a Volterra type
integral equation involving the biscalar $K$ (see Appendix B for details)
$$ \lambda_\eps(x) + {1\over 2\pi} \int_{C^-_\eps(x)}\!\!\!
   \boxeps K_\eps(\xi,x) \lambda_\eps(\xi) 
   \,\mueps_\Gamma(\xi) 
= {1\over 2\pi} \int_{C^-_\eps(x)}\!\!\!
   K_\eps(\xi,x) \psi(\xi) \,\mueps_{\Gamma}(\xi)
   \tag\tagnum\xlabel{volterra} $$   
where $\mueps_{\Gamma}$ is the volume element induced by $\mueps$ on
$C^-_\eps(x)$.  We next examine the limiting behaviour of the terms
in~\eqref{volterra}. We first consider the integral on the left hand
side. In Appendix B it is shown that in normal coordinates based at $\xi$
$$
\boxeps K_\eps(\xi,x)=-{1 \over 6}K_\eps(\xi,x)R_\eps(\xi)+O(\xi^2)
\tag\tagnum\xlabel{eq1}
$$
On the other hand using the results of Clarke et al (1996)
$$
\lim_{\eps \to 0}\int
R_\eps(\xi)\Psi(\xi)\mu(\xi)=4\pi(1-A)\int\Psi(\xi^0,0,0,\xi^3)
d\xi^0d\xi^3
\tag\tagnum\xlabel{eq2}
$$
So that provided $\lambda_0$ is well defined then
$$
\lim_{\eps \to 0}
\int_{C^-_\eps(x)}\!\!\!\!\!
   \boxeps K_\eps(\xi,x) \,\lambda_\eps(\xi) 
   \,\mueps_\Gamma(\xi) = 
  -{2 \over 3}(1-A)\pi
 \int_{\Lambda^-(x)}\!\!\!\!\!\lambda_0(\xi)\,\mu_{\Lambda^-}(\xi)
\tag\tagnum\xlabel{volterra1}
$$   
where $\Lambda^-(x)$ is the limit of the intersection of $C^-_\eps(x)$
with the axis $\Lambda$ and $\mu_{\Lambda^-}$ is the volume form induced
on $\Lambda^-(x)$. It is shown in Appendix C that these quantities are
well defined.

The integral of the right hand side of~\eqref{volterra} is more
straightforward since for $x, \  \xi \notin \Lambda$, $K_\eps(\xi,x)
\to 1$ as $\eps \to 0$. Also we show in Appendix C that $C^-_\eps(x)$
tends to a well defined limit $C^-_0(x)$. However some care must be
taken in interpreting $C^-_0(x)$. One cannot simply take this to be
the past null cone of $x$ in the conical space-time with the axis
removed. This is because deleting the axis results in a tear in the
null cone and destroys the $S^2$ topology. However (in the case of a
rotationally symmetric smoothing) it is shown in Appendix C that 
$C^-_0(x)=\lim_{\eps \to 0}C^-_\eps(x)$ is well defined and does have
$S^2$ topology the missing piece being generated by geodesics which
pass within $O(\eps)$ of the axis $\Lambda$, and which hit the axis in
the limit. One can also show that the corresponding volume form
$\mu^\eps_\Gamma(\xi)$ also has a well defined limit for $\xi \notin \lambda$
given by the volume form induced by the conical metric on
$C^-_0(x)$. Furthermore the value on the axis is bounded so that
we need not include the contribution from the integral over the axis.
Hence provided that $\lambda_0$ exists we have
$$
\lim_{\eps \to 0}
\int_{C^-_\eps(x)}\!\!\!\!\!
   K_\eps(\xi,x)\, \psi(\xi) \,\mueps_{\Gamma}(\xi)
=\int_{\hat C^-_0(x)}\!\!\!\!\! \psi(\xi)\,\mu^0_\Gamma(\xi)
\tag\tagnum\xlabel{volterra2} $$   
where $\hat C^-_0(x)=C^-_0(x) \setminus \Lambda$.

On the basis of~\eqref{volterra1} and~\eqref{volterra2} we now {\it define}
$\lambda(x)$, to be the solution of
$$
  \lambda(x)-{1 \over 3}(1-A)\int_{\Lambda^-(x)}\!\!\!\!\!
  \lambda_0(\xi)\,\mu_\Lambda(\xi)
  ={1 \over {2\pi}}\int_{\hat C^-_0(x)}\!\!\!\!\!
  \psi(\xi)\,\mu^0_\Gamma(\xi) 
  \tag\tagnum\xlabel{voltlim} $$   

Note that this gives $\lambda$ in terms of $\psi$ on $\hat C^-_0(x)$ and
$\lambda$ on $C^-_0(x) \cap \Lambda$. For a point $x$ that lies on the
axis $\Lambda^-(x)$ degenerates into a pair of null lines and
$\mu_\Lambda$ vanishes so that the integral on the left hand side 
of~\eqref{voltlim} vanishes and we have
$$
\lambda(x)={1 \over {2\pi}}\int_{\hat C^-_0(x)}\!\!\!\!\! \psi(\xi)\,
\mu^0_\Gamma(\xi),
\qquad x \in \Lambda
\tag\tagnum\xlabel{voltlim2} $$   
where for a point $x$ on the axis, the past null cone $\hat C^-_0(x)$
is given in quasi-Cartesian coordinates by the usual Minkowskian formula. 
Equation~\eqref{voltlim2} gives an expression for $\lambda \in \Lambda$
simply in terms of an integral involving $\psi$, so we may substitute
back for $\lambda$ in the left hand integral of~\eqref{voltlim} to
obtain an integral for $\lambda$ in terms of $\psi$ which is valid for
all $x$.
$$
\lambda(x)=
{1 \over {2\pi}}\int_{\xi \in C^-_0(x)}\!\!\!\!\!
\psi(\xi)\,\mu^0_\Gamma(\xi)
+{1 \over {6\pi}}(1-A)\int_{\xi \in \Lambda^-(x)}\!\biggl(
\int_{\eta \in C^-_0(\xi)}\!\!\!\!\!\psi(\eta)\,
\mu^0_{\Gamma(\xi)}(\eta)\biggr)
\mu_\Lambda(\xi)
\tag\tagnum\xlabel{lambda} $$   
Thus a solution to~\eqref{voltlim} is given by the above 
integral~\eqref{lambda}. Clearly this is a well defined quantity which 
does not depend on the choice of the smoothing kernel $\varphi$.
It remains to show that $\lambda_\eps$ does indeed tend to $\lambda$
as $\eps \to 0$. To do this we let
$\rho_\eps(x)=\lambda(x)-\lambda_\eps(x)$
then subtracting~\eqref{voltlim} from~\eqref{volterra} we find that
$\rho_\eps$ satisfies the integral equation
$$
\aligned
\rho_\eps(x)+
{1\over 2\pi} \int_{C^-_\eps(x)}\!\!\!\!\!
   \boxeps K_\eps(\xi,x)\, \rho_\eps(\xi) 
   \,\mueps_\Gamma(\xi) 
&={1\over 2\pi} \int_{C^-_\eps(x)}\!\!\!\!\!
   \boxeps K_\eps(\xi,x)\, \lambda(\xi) 
   \,\mueps_\Gamma(\xi) +
{1 \over 3}(1-A)\int_{\Lambda^-(x)}\!\!\!\!\!
\lambda_0(\xi)\,\mu_\Lambda(\xi)\\
&+{1 \over {2\pi}}\int_{\hat C^-_0(x)}\!\!\!\!\!
\psi(\xi)\,\mu^0_\Gamma(\xi)
-{1 \over {2\pi}}\,\int_{C^-_\eps(x)}\!\!\!\!\!
\psi(\xi)\,\mu^\eps_\Gamma(\xi)  \\
\endaligned
\tag\tagnum\xlabel{rho} 
$$   
where $\lambda$ is given by~\eqref{lambda}.

Since we know from~\eqref{lambda} that $\lambda$ is bounded then~\eqref{eq1}
and~\eqref{eq2} show that
$$
{1\over 2\pi} \int_{C^-_\eps(x)}\!\!\!\!\!
   \boxeps K_\eps(\xi,x) \lambda(\xi) 
   \,\mueps_\Gamma(\xi) +
{1 \over 3}(1-A)\int_{\Lambda^-(x)}\!\!\!\!\!
\lambda_0(\xi)\mu_\Lambda(\xi)
=O(\eps)
$$
Also from Appendix C we know that $C^-_\eps(x) \to C^-_0(x)$ as $\eps
\to 0$, and that $\mu^\eps_\Gamma \to \mu^0_\Gamma$, except possibly
on $\Lambda$ (which is of measure zero) so that
$$
{1 \over {2\pi}}\int_{\hat C^-_0(x)}\!\!\!\!\! \psi(\xi)\,\mu^0_\Gamma(\xi) 
-{1 \over {2\pi}}\int_{C^-_\eps(x)}\!\!\!\!\!
\psi(\xi)\,\mu^\eps_\Gamma(\xi) =O(\eps)
$$

Hence $\rho_\eps(x)$ satisfies the integral equation
$$
\rho_\eps(x)+
{1\over 2\pi} \int_{C^-_\eps(x)}\!\!\!\!\!
   \boxeps K_\eps(\xi,x)\, \rho_\eps(\xi) 
   \,\mueps_\Gamma(\xi) 
=h_\eps(x)
\tag\tagnum\xlabel{rho2} 
$$
where $h_\eps(x)$ denotes the function on the right hand side
of~\eqref{rho} and which tends to zero as $\eps \to 0$.
We may obtain a solution to this equation by iteration and find that
$\rho_\eps(x)$ also tends to zero as $\eps \to 0$.

Thus $\lambda_\eps(x)$ tends to $\lambda(x)$ given by~\eqref{lambda}
as $\eps \to 0$ and inserting this into~\eqref{intupsi} and using~\eqref{lim}
we see that for $\varphi \in \cA_q$ for $q$ sufficiently large
$$
\lim_{\eps \to 0}\langle u_\eps, \psi \rangle=
-\int_{S\setminus \Lambda}\!\!\bigl( \partial_t \lambda(0,x^\alpha)
   v(x^\alpha) + \lambda(0,x^\alpha) w(x^\alpha)
   \bigr)\, \mu^0_S(x^\alpha) 
\tag\tagnum\xlabel{answer}
$$
where $\mu^0_S$ is the volume form induced by the conical metric on $S
\setminus \Lambda$ and $\lambda(t,x^\alpha)$ is given
by~\eqref{lambda}.  Thus $U$ is associated to a distribution $u$ defined
by~\eqref{lambda} and the right hand side of~\eqref{answer}.

It is worth remarking that even though $M \setminus \Lambda$ is locally
flat the Greens function for the cone is not sharp due to the second term
in~\eqref{lambda} (which vanishes in Minkowski space when $A=1$).  The
solution at $x$ depends not just on the initial data on $C^-_0(x)
\cap S$ but also on points in $C^-_0(\xi) \cap S$ with $\xi \in
C^-_0(x) \cap \Lambda$. i.e.\ as well as the sharp term in the Greens
function there is an extra term which involves scattering off the axis.

\head
6. Conclusion
\endhead

In an earlier paper (Clarke et al 1996) we showed that it was possible
to give a distributional interpretation to the curvature of a conical
space-time by using Colombeau's theory of non-linear generalised
functions. In this paper we have established that such conical
singularities do not disrupt the Cauchy development of test fields on
this fixed background. Although it is not feasible to undertake a full
non-linear analysis and show that such singularities do not disrupt
the Cauchy development of Einstein's equations, the higher order
energy estimates obtained in Section 4 indicate that the back reaction
is not likely to radically affect the nature of the singularity. These
results taken together therefore show that it is reasonable to
interpret conical space-times not only as distributional geometries
but also as distributional solutions of Einstein's equations.

The concept of generalised hyperbolicity considered in this paper
($\cG$-hyperbolicity) is different from that adopted by Clarke
(1998) when considering curve-integrable space-times. However the
curve integrable condition is primarily used to construct a geodesic
congruence with tangent vector having bounded covariant derivative,
and such a congruence may be explicitly constructed in conical
space-times even though they fail to be curve-integrable. In a future
paper we will examine the relationship between these two conditions.

Finally it is worth remarking that the limiting solution that one
obtains to the wave equation is more interesting than one might at
first expect. It contains a term due to the delta-function curvature
on the axis and therefore has tail terms even though the conical
space-time $(M \setminus \Lambda, g_0)$ is locally flat. It also
contains a term which arises from parts of the null cone which in the
limit pass through the axis and `mends the tear'. A naive approach in
which one looked at the solution in Minkowski space in cylindrical
polar coordinates and then rescaled the angular coordinate would not
produce the correct answer.

\eqcount=0
\def\eqprint{A.\number\eqcount}
\head
Appendix A. Restrictions in the Colombeau algebra
\endhead

When dealing with initial value or boundary value problems in the
context of Colombeau algebras one is interested in solving some
differential equation on $\Real^n$, but giving data on some lower
dimensional subspace $S \equiv \Real^m$. For example when solving the
initial value problem for the wave equation in $\cG(\Real^4)$ 
$$\aligned
\square U &=0 \\
U_{|S}&=V \\
\partial_t U_{|S}&=W \\
\endaligned
$$
where $S=\{(t,x,y,z) \in \Real^4 : t=0 \} \equiv \Real^3$, we need to
regard $U \in \cG(\Real^4)$ but $U_{|S} \in \cG(\Real^3)$. One way of
doing this is to require that elements $\varphi^{(n)} \in
\cA(\Real^n)$ are products of one dimensional kernels $\varphi \in 
\cA(\Real)$ so that
$$
\varphi^{(n)}(x^1,\dots,x^n)=\varphi(x^1)\cdots \varphi(x^n)
$$
If $U \in \cG(\Real^4)$ then we may define $U_{|S} \in \cG(\Real^3)$
by
$$
U_{|S}(\varphi^{(3)}, (x,y,z))=U(\varphi^{(4)},(0,x,y,z))
\tag\tagnum\xlabel{cartesian}
$$
Unfortunately we often want to restrict $\cA_q(\Real^4)$ to some
subset which is invariant under the action of some symmetry group (for
example rotations about the $z$-axis) and it is not usually possible
to achieve this with kernels taking the form~\eqref{cartesian}. In
this appendix we will therefore describe a way of defining
restrictions which allows one to work with kernels with a specified
symmetry. We will illustrate the approach by restricting
$\cG(\Real^4)$ to a three dimensional subspace, but the construction
can be readily generalised to any linear subspace of $\Real^n$.

Let $\tilde f \in \cE_M(\Real^4)$, then we define
$$\aligned
\tilde f_{|S}:\cA_q(\Real^3) \times \Real^3 &\to \Real \\
\hbox{by}\quad \tilde f_{|S}(\varphi,(x,y,z))&=\tilde f(\hat\varphi, (0,x,y,z))
\\ \endaligned
$$
where
$$
\hat\varphi(t,x,y,z)=\tfrac13\varphi(x,y,z)\int_{\Real^2}\!
\bigl(\varphi(t,u,v)+\varphi(v,t,u)+\varphi(u,v,t)\bigr)\,du\,dv
$$
The first important point to note is that if $\varphi \in \cA(\Real^3)$ is
invariant under some symmetry group (such as rotations about the
$z$-axis) then $\hat\varphi \in \cA(\Real^4)$ is invariant under the
same symmetry. The second important feature of the construction is
given by the following Lemma.

\proclaim{Lemma \procnum}
\xlabel{Aq}
\roster
\item
$\quad \varphi \in \cA_0(\Real^3) \implies \hat\varphi \in \cA_0(\Real^4)$
\item
$\quad \varphi \in \cA_q(\Real^3) \implies \hat\varphi \in \cA_q(\Real^4)$
\endroster
\endproclaim

\demo{Proof}
For (1) we note
$$\aligned
\int_{\Real^4}\!\hat\varphi(t,x,y,z)\,dt\,dx\,dy\,dz
&=\tfrac13\int_{\Real^3}\!\varphi(x,y,z)\,dx\,dy\,dz
\int_{\Real^3}\!\bigl(\varphi(t,u,v)+\varphi(v,t,u)+\varphi(u,v,t)\bigr)
\,du\,dv\,dt \\
&=1
\endaligned
$$
For (2) we let $\varphi \in \cA_q(\Real^3)$ with $q \geq 1$ and let $a,\,
b,\,c,\,d\in\Nat$ be such that $1 \leq a+b+c+d \leq q$. Then
$$
\multline
 \int_{\Real^4}\!\hat\varphi(t,x,y,z)t^a x^b y^c z^d\,dt\,dx\,dy\,dz\\
=\tfrac13\int_{\Real^3}\!\varphi(x,y,z)x^b y^c z^d\,dx\,dy\,dz
\!\int_{\Real^3}\!\bigl(\varphi(t,u,v)+\varphi(v,t,u)
+\varphi(u,v,t)\bigr)t^a\,du\,dv\,dt 
\endmultline
$$
If $a \geq 1$, the second integral on the right vanishes, while if
$a=0$ the first integral on the right vanishes. In either case the
left hand side vanishes and $\hat\varphi \in \cA_q(\Real^4)$.
\enddemo

\proclaim{Proposition \procnum}
\xlabel{mod}
\roster
\item
$\quad \tilde f \in \cE_M(\Real^4) \implies \tilde f_{|S} \in \cE_M(\Real^3)$
\item
$\quad \tilde f \in \cN(\Real^4) \implies \tilde f_{|S} \in \cN(\Real^3)$
\endroster
\endproclaim

\demo{Proof}
The proof of (1) and (2) follows directly from the definition of
moderate and null together with Lemma~\xref{Aq}.
\enddemo

\proclaim{Corollary \procnum}
\xlabel{corol}
Let $F \in \cG(\Real^4)$ have representative $\tilde f \in
\cE_M(\Real^4)$, then $F_{|S} \in \cG(\Real^3)$ may be defined by 
$F_{|S}=[\tilde f_{|S}]$.
\endproclaim

We have therefore shown how it is possible to restrict elements of
$\cG(\Real^4)$ to some subspace $S$ while maintaining the required
symmetry. The following result shows that for smooth functions this
notion of restriction commutes with the canonical embedding
$\iota_n:C^\infty(\Real^n)\to\cG(\Real^n)$.
\proclaim{Proposition \procnum}
\xlabel{commutes}
Let $f \in C^\infty(\Real^4)$ then
$$
(\iota_4 f)_{|S}=\iota_3(f_{|S})
$$
\endproclaim
\demo{Proof}
$$\aligned
\widetilde{(f_{|S})}_\eps&=\int_{\Real^3}\!f(0,x+\eps x',y+\eps y',z+\eps z')
\varphi(x',y',z')\,dx'\,dy'\,dz' \\
(\tilde f_\eps)_{|S}&=
\!\tfrac13\int_{\Real^6}\!f(\eps t',x+\eps x', y+\eps y', z+\eps z')
\varphi(x',y',z')\, \\
&\qquad\qquad\qquad\qquad\times\bigl(\varphi(t',u,v)
+\varphi(v,t',u)+\varphi(u,v,t')\bigr)\,du\,dv\,dt'\,dx'\,dy'\,dz'
\\ 
\endaligned
$$
Then by looking at the Taylor expansion of
$\widetilde{(f_{|S})}_\eps-(\tilde f_\eps)_{|S}$ one can verify that
the difference is null.
\enddemo

\eqcount=0
\def\eqprint{B.\number\eqcount}

\head
Appendix B. The Greens function for the wave operator $\square$.
\endhead

In this appendix we summarise some results of Friedlander (1975) and Bruhat
(1962).  Let $x\in M$. A distribution may be defined on $D^-(x)$ as
follows; let
$$ S^-_\eps(x) = \{ \xi\in D^-(x) | \Gamma(\xi,x)=\eps \} $$
where $\Gamma(\xi,x)$ denotes the square distance between $x$ and $\xi$ in
$M$. We then define a distribution $\delta^-(\Gamma(\xi,x)-\eps)$ by
$$ \ip{\delta^-(\Gamma(\xi,x)-\eps)}{\varphi(\xi)} = \int_{S^-_\eps(x)}
   \!\!\!\!\! \varphi(\xi)\, \mu_\Gamma(\xi) $$
Consequently we may define a limiting distribution $\delta^-(\xi,x)$ by
$$ \ip{\delta^-(\xi,x)}{\varphi(\xi)} = \lim_{\eps\to0^+} \int_{S^-_\eps(x)}
   \!\!\!\!\! \varphi(\xi)\, \mu_\Gamma(\xi) $$

It may be shown, for the wave operator $\square$ (Friedlander 1975, Theorem
4.2.1), that there exists a function $K\in C^\infty \times C^\infty$ (the
biscalar) such that
$$ \square(K(\xi,x)\delta^-(\xi,x)) = (\square K(\xi,x))\delta^-(\xi,x) +
   2\pi\delta(\xi-x) \tag\tagnum\xlabel{friedlander}$$
and moreover $K$ can be chosen so that
$$
\alignat1
   &2 \nabla{}^a\Gamma \nabla{}_a K + ( \square\Gamma-8)K = 0
     \tag\tagnum\xlabel{KGamma}\\
   &K(x,x) = 1 
\endalignat
$$
On integrating~\eqref{friedlander} we obtain
$$ \int_M\! \square(K(\xi,x)\delta^-(\xi,x)) \varphi(\xi) \mu(\xi) = \int_M 
   \!(\square K(\xi,x))\delta^-(\xi,x) \varphi(\xi) \mu + 2\pi\varphi(x) $$
which may be rearranged, by integrating by parts and using the fact that
$\varphi$ has compact support, to obtain
$$ \varphi(x) = {1\over 2\pi} \int_{C^-(x)} \!\!\!\!\! K(\xi,x)\, \square
   \varphi(\xi) \,\mu_{\Gamma}(\xi) - {1\over 2\pi} \int_{C^-(x)}
   \!\!\!\!\!\square K(\xi,x)\, 
   \varphi(\xi)\,\mu_{\Gamma}(\xi) $$
This equation may be regarded as a Volterra type integral equation for
$\varphi$ given $\psi\in\cD$ ($\square\varphi$ in this case)
$$ \varphi(x) + {1\over 2\pi} \int_{C^-(x)}\!\!\!\!\! \square K(\xi,x)\,
   \varphi(\xi)\,\mu_{\Gamma}(\xi) = {1\over 2\pi} \int_{C^-(x)}\!\!\!
   K(\xi,x) \,\psi(\xi) \,\mu_{\Gamma}(\xi) $$ 
This equation naturally defines a distribution $\green-{x}:\cD\to\Real$
given by $\ip{\green-{x}}{\psi}=\varphi(x)$ (Bruhat, 1962). The distribution
$\green-{x}$ is often known as the past Greens function.  A future Greens
function $\green+{x}$ may be defined by an analogous procedure, replacing
the past null cone $C^-(x)$ with the future null cone $C^+(x)$ in all the
definitions above.  In fact we may define a distributions on
$\cD\otimes\cD$ by
$$ \ip{G^{\pm}(x,\xi)}{\varphi(x)\psi(\xi)} =
   \ip{\ip{E^\pm_x}{\psi}}{\varphi(x)} $$ 
The fact that $\square$ is self adjoint implies that $G^+(x,\xi)=G^-(\xi,x)$.

\subhead
Calculation of $K$ and $\square K$ for the wave operator $\square$
\endsubhead

We shall first consider $K(x,y)$ in normal coordinates based at $x$. In
normal coordinates we have
$$
\aligned
   &g_{ab}(y) y^b \normeq g_{ab}(x) y^b \\
   &\Gamma(x,y) \normeq g_{ab}(x) y^a y^b
\endaligned
$$ 
This implies that
$$   \nabla^a\Gamma \normeq 2 y^a  $$
and so
$$
\split \square \Gamma
   &\normeq |g|^{-1/2} \partial_a \bigl( |g|^{1/2} \nabla^a \Gamma\bigr) \\
   &\normeq 2|g|^{-1/2} \partial_a \bigl( |g|^{1/2} y^a \bigr) \\
   &\normeq 8+ y^a \partial_a \bigl( \log |g| \bigr)
\endsplit
$$
Substituting in~\eqref{KGamma} gives
$$ 4 y^a \partial_a K + y^a \partial_a \bigl( \log |g| \bigr) K \normeq 0 $$
which implies
$$ y^a \partial_a(|g|^{1/4}) \normeq 0 $$
This means that $|g(y)|^{1/4} K(x,y)$ is constant along the geodesic
connecting $x$ and $y$. Therefore
$$
\split
  |g(y)|^{1/4} K(x,y) & \normeq |g(x)|^{1/4} K(x,x) \\
  & \normeq |g(x)|^{1/4}
\endsplit
$$
Therefore
$$ K(x,y) \normeq {|g(x)|^{1/4} \over |g(y)|^{1/4} }
   \tag\tagnum\xlabel{Knormal}$$ 

We now consider $K(x,y)$ in a more general coordinate system. We shall use
primed coordinates to denote the normal coordinate system.  Let
$$ J(y)=\det\biggl(\pbyp{y^a}{y'{}^b}\biggr) $$
Then
$$
\split  K(x,y)
  &= {|g'(x)|^{1/4} \over |g'(y)|^{1/4}} \\
  &= {|g(x)|^{1/4}{J(x)}^{-1/2} \over |g(y)|^{1/4} J(y)^{-1/2}} \\
  &= {|g(x)|^{1/4} \over |g(y)|^{1/4}} J(y)^{1/2}
\endsplit
$$
But
$$ \pbyp{\Gamma}{{x}^a}(x,y) = -2 g_{ab}(x) y'{}^b $$
So
$$ \pbypp{\Gamma}{y^a}{x{}^b}(x,y) = -2 g_{bc}(x)
   \pbyp{y'{}^c}{y^a}  $$
implying that
$$ \det \biggl( \pbypp{\Gamma}{y^a}{x{}^b}(x,y) \biggr) = -2
   g(x) J(y)^{-1} $$
Therefore
$$ K(x,y) = {\biggl| \det \bigl( \pbypp{\Gamma}{y^a}{x{}^b}
   \bigr) \biggr|^{1/2} \over 4 \bigl| g(x) \bigl|^{1/4} \bigl| g(y)
   \bigl|^{1/4} } $$

We shall now consider $\square K(x,y)$ in normal coordinates based at
$x$. Applying $\square$ to ~\eqref{Knormal}
$$
\split
  \square K &= |g|^{-1/2} \partial_a ( g^{ab} |g|^{1/2} \partial_b K)\\
  &= |g|^{-1/4} \partial_a \bigl(g^{ab} |g|^{1/2} \partial_b(|g|^{-1/4})
  \bigr) \\
  &= -\tfrac14 K(x,y) \gamma(y)
\endsplit
$$
where
$$ \gamma(y) \normeq (g^{ab} g^{cd} g_{cd,a})_b + \tfrac14 g^{ab} g^{cd}
g^{ef} g_{cd,a} g_{ef,b} $$

In normal coordinates at $x$ it may be shown that
$$
\split
   &g_{ab,c}(x) \normeq 0 \\
   &R(x) \normeq \tfrac32 g^{ab}(x) g^{cd}(x) g_{cd,ab}(x)
\endsplit
$$
so
$$\gamma(x) \normeq \tfrac23 R(x) $$
We may now expand $g_{ab}(y)$ in powers of $y$ about $x$;
$$ g_{ab}(y) \normeq g_{ab}(x) + \tfrac13 R_{acbd}(x) y^c y^d + O(y^3) $$ 
this implies
$$ \gamma(y) \normeq \tfrac23 R(x) + \tfrac13 \bigl( \nabla_a R(x) + 2
\nabla^b R_{ab}(x) \bigr) y^a + O(y^2) $$
and on applying the Bianchi identities,
$$ \gamma(y)-\tfrac23 R(x) \normeq O(y^2) $$
Therefore
$$ \square K(x,y) \normeq -\tfrac16 K(x,y) R(y) + O(y^2)
   \tag\tagnum\xlabel{boxK} $$
a result obtained by McLenaghan (1969).

\eqcount=0
\def\eqprint{C.\number\eqcount}

\head
Appendix C: Null geodesics in a conical space-time.
\endhead

In this appendix we examine the limit of $C^-_\eps(x)$, the past null
cone of $x$ in $(M,g_\eps)$ as $\eps \to 0$. To do this we first
examine the past null cone of points in  $(M\setminus \Lambda, g_0)$,
the conical space-time with the axis removed. Although it is not hard
to derive the equation of the null cone by working in the Minkowskian 
covering space and then making the appropriate identifications, we will
derive it using a method that can also be applied to 
$(M\setminus \Lambda, g_\eps)$.

If we work in polar coordinates the geodesics may be found using the
Lagrangian 
$$
L=\tfrac12\{-\dot t^2+\dot r^2 +A^2r^2\dot \phi^2 +\dot z^2\}
$$
where a dot indicates the derivative with respect to some affine
parameter. The geodesic equations show that we may take $t$ as an
affine parameter. In view of the rotational and translational
symmetries of the metric, without loss of generality we may consider the
null cone of a point $P$ with coordinates $(0,a,0,0)$. We therefore
consider geodesics which, when $t=0$, satisfy the initial conditions:
$$\xalignat{3}
   r&=a & \phi&=0 & z&=0 \\
   \dot{r}&=-\sin\gamma\cos\beta &
   \dot{\phi} &= {1\over Aa} \sin\beta\sin\gamma &
   \dot{z} &=\cos\gamma \\
\endxalignat
$$
where $\beta$ and $\gamma$ are polar angles which parameterise the
$S^2$ celestial sphere of null directions at $P$ (with $\beta=0$
corresponding to a geodesic which hits the axis). 

Then the $z$ equation gives 
$$
z=t\cos\gamma
$$
The null geodesics may then be found by solving the equations
$$\aligned
A^2r^2\dot\phi&=J \\
\dot r^2+A^2r^2\dot\phi^2&=E^2
\endaligned
$$
where
$$\aligned
J&=Aa\sin\beta\sin\gamma \\
E^2&=\sin^2\gamma
\endaligned
$$
Substituting for $\dot \phi$ this gives
$$
{{dr}\over {dt}}=\pm{{\sin\gamma(r^2-a^2\sin^2\beta)^{1/2}} \over r}
$$
and hence
$$
t=-\int {{rdr} \over {\sin\gamma(r^2-a^2\sin^2\beta)^{1/2}}}
$$
where the minus sign is taken for the past null cone.
Performing the integration and using the fact that $r=a$ when $t=0$
gives
$$
t= -{{(r^2-a^2\sin^2\beta)^{1/2}}\over {\sin\gamma}}+
{{a\cos\beta} \over {\sin\gamma}}
$$
Solving for $r$ enables one to find $r(t)$ where
$$
r^2(t)=t^2\sin^2\gamma-2at\cos\beta\sin\gamma+a^2
\tag\tagnum\xlabel{r}
$$
We now substitute for $r$ in the $\phi$ equation to obtain
$$
{{d\phi}\over {dt}}={{a\sin\beta\sin\gamma}\over
{A(t^2\sin^2\gamma-2at\cos\beta\sin\gamma+a^2)}}
$$
Integrating and putting in the initial condition that $\phi=0$ when
$t=0$ gives
$$
A\phi=-\beta\pm\cot^{-1}\left({{t\sin\gamma-a\cos\beta}\over {a\sin\beta}}\right)
\tag\tagnum\xlabel{phi}
$$
We have now obtained $r(t)$, $\phi(t)$ and $z(t)$ and one can now
eliminate $\beta$ and $\gamma$ and show that $t,\ r,\ \phi$ and $z$ satisfy 
the constraint
$$
a^2+r^2-2ar\cos(A\phi)+z^2-t^2=0
\tag\tagnum\xlabel{constraint}
$$
The points on  the past null cone of $(0,a,0,0)$ are therefore the
points where $t<0$ and~\eqref{constraint} is satisfied.

We now turn to considering the null cone in the regularised
space-time $(M,g_\eps)$. If we use a rotationally symmetric smoothing
kernel, as described in Appendix A, we may write the metric in the
form
$$ 
ds^2=-dt^2+P^2_\epsilon(r) \,dr^2 + r^2 Q^2_\epsilon(r) \,d\phi^2 
   + dz^2 $$ 
where $P_\eps(r)=P(r/\eps)$, $Q_\eps(r)=Q(r/\eps)$ and $P$ and $Q$
are given by
$$ \eqalign{
   P(r)^2 &= \tfrac12(1+A^2)+\tfrac12(1-A^2) 2\pi\int^r_0
   \Bigl(1-{\rho^2\over r^2} \Bigr) \varphi(\rho)\rho\,d\rho \cr
   Q(r)^2 &= \tfrac12(1+A^2)-\tfrac12(1-A^2) 2\pi\int^r_0
   \Bigl(1-{\rho^2\over r^2} \Bigr) \varphi(\rho)\rho\,d\rho \cr
   }$$
Here $\varphi$ is a smooth function with compact support such that
$$ \eqalign{
   & 2\pi\int\varphi(r) r\,dr=1 \cr
   & 2\pi\int\varphi(r) r^3\,dr=0 \cr
}$$
so that the metric (when regarded in Cartesian coordinates as a function of
$\varphi$) may be interpreted as an element of Colombeau's generalised
function algebra ${\cal G}(\Real^4)$.

In this space-time the geodesics may be found by considering the
Lagrangian
$$
L_\eps=\tfrac12\{-\dot t^2+P^2_\eps\dot r^2 +r^2Q^2_\eps\dot \phi^2 
+\dot z^2\}
$$
and we see that as before we may take $t$ as an affine parameter. We
again consider the null cone of $(0,a,0,0)$, but this time take the
initial conditions:
$$\xalignat3
   r&=a & \phi&=0 & z&=0 \\
   \dot{r}&=-{1 \over {P_\eps(a)}}\sin\gamma\cos\beta &
   \dot{\phi} &= {1\over {aQ_\eps(a)}} \sin\beta\sin\gamma &
   \dot{z} &=\cos\gamma \\
\endxalignat
$$

We again have $z(t)=t\cos\gamma$, and the $r$ and $t$ equations are now
$$\aligned
Q_\eps^2(r)r^2\dot\phi&=J_\eps \\
P^2_\eps(r)\dot r^2+r^2Q^2_\eps(r)\dot\phi^2&=E_\eps^2
\endaligned
$$
where
$$\aligned
J_\eps&=aQ_\eps(a)\sin\beta\sin\gamma \\
E_\eps^2&=\sin^2\gamma
\endaligned
$$
Substituting for $\dot \phi$ this gives
$$
{{dr}\over
 {dt}}=\pm{{\sin\gamma(Q^2_\eps(r)r^2-a^2Q^2_\eps(a)\sin^2\beta)^{1/2}}
 \over {P_\eps(r)Q_\eps(r)r}}
$$
and hence
$$
t=-\int {{P_\eps(r)rdr} \over {\sin\gamma(r^2-a^2S_\eps(r)\sin^\beta)^{1/2}}}
$$
where $S_\eps(r)=Q_\eps(a)/Q_\eps(r)$.

We now let
$$
R_0=\sup\{r:|\varphi(r)| >0\}
$$
and for the moment restrict attention to geodesics for which
$\rmin>\eps R_0$. Then for $\varphi \in \cA_q$ and $r >\eps$ 
we have
$$\aligned
P_\eps&=1+O\left({{\eps^{q+1}} \over {r^{q+1}}}\right)
R_0) \\ 
Q_\eps&=A+O\left({{\eps^{q+1}} \over {r^{q+1}}}\right)
R_0) \\ 
\endaligned
$$
These estimates allow us to deduce that
$$
t= -{{(r^2-a^2\sin^2\beta)^{1/2}}\over {\sin\gamma}}+
{{a\cos\beta} \over {\sin\gamma}}+g_\eps(r)
$$
where $g_\eps(r) \to 0$ uniformly as $\eps \to 0$, and hence that
$$
r_\eps(t)= r(t)+h_\eps(t)
\tag\tagnum\xlabel{reps}
$$
where $r(t)$ is given by~\eqref{r} and $h_\eps(t) \to 0$ uniformly as
$\eps \to 0$.

We now consider the $\phi$ equation
$$
{{d\phi} \over {dt}} ={{J_\eps} \over {Q^2_\eps(r)r^2}}
$$
so that
$$
Q_\eps(a)\phi=\int {{aS^2_\eps(r)\sin\beta\sin\gamma\, dt} \over
{t^2\sin^2\gamma-2at\cos\beta\sin\gamma+a^2+h_\eps(t)}}
$$
Using the initial conditions for $\phi$ and fact that $r_\eps(t) >\eps
R_0$ together with  the   estimates for  $P_\eps$ and $Q_\eps$ we may 
deduce that
$$
\phi_\eps(t)=\phi(t)+k_\eps(t)
\tag\tagnum\xlabel{phieps}
$$
where $\phi(t)$ is given by~\eqref{phi} and $k_\eps(t) \to 0$
uniformly as $\eps \to 0$.

If we now substitute~\eqref{reps} and~\eqref{phieps} into the
constraint equation~\eqref{constraint} for the null cone in a conical 
space-time we find we get an equation of the form
$$
a^2+r^2-2ar\cos(A\phi)+z^2-t^2=m_\eps(t,r,\phi,z)
$$
where $m_\eps \to 0$ uniformly as $\eps \to 0$.

We also note that as $\eps \to 0$, the excluded geodesics are those which
hit the axis. So that if we denote by $\hat C^-(x)$ the past null cone of
$x$ excluding those points which lie on geodesics which hit the axis, then
the above result implies that $\hat C^-_\eps(x) \to \hat C^-_0(x)$ as
$\eps\to0$.

The estimates for $P_\eps$ and $Q_\eps$ also show that the volume form
$\mu^\eps$ defined by $g_\eps$ tends uniformly to a well defined limit
$\mu^0$ as $\eps \to 0$ (at least for $x \notin \Lambda$). So that if
$\psi \in \cD(M)$ then
$$
\lim_{\eps\to0}\int_{\hat C^-_\eps(x)}\!\!\!\!\!
\psi(\xi)\,\hat\mu_\Gamma^\eps(\xi) 
= \int_{\hat C^-_0(x)}\!\!\!\!\!\psi(\xi)\,\hat\mu_\Gamma^0(\xi)
\tag\tagnum\xlabel{intlim}
$$
where $\hat \mu^\eps_\Gamma(\xi)$ and $\hat \mu^0_\Gamma(\xi)$ are the volume
forms on $\hat C^-_\eps(x)$ and $\hat C^-_0(x)$ induced by $\mu^\eps$
and $\mu^0$ respectively.

However, as we have already observed $\hat C^-_0(x)$ is {\it not} the
limit of $C^-_\eps(x)$ since it does not contain the part generated by
geodesics which get closer to the axis than $\eps R_0$, 
and therefore contains a
tear which destroys the $S^2$ topology. We now consider the portion of
the null cone generated by such geodesics (which restores the $S^2$ topology).
For a fixed value of $\gamma$ the geodesics which generate
the missing piece are given by those which satisfy the initial
condition that  $\beta$ is proportional to $\eps$.
Since the geodesics which satisfy $\rmin >\eps R_0$ tend to straight
lines as $\eps \to 0$, when written in quasi-Cartesian coordinates,
the previously excluded geodesics must
satisfy $\tan \beta <\eps R_0/a$ (for sufficiently small
$\eps$). We therefore consider the surface $S_\eps(P)$
generated by past directed null geodesics $\kappa^\eps_{s,\gamma}(t)$ 
which emanate from the point $P$ (which we may assume has coordinates 
$(0,a,0,0)$) making polar angles
$$
\gamma \in (0,\pi), \qquad \beta=s\eps R_0/a \quad s \in [-1,1]
$$
and  then consider the limit of this surface as $\eps \to 0$.

Let $D_\eps$ be the region for which $r<\eps R_0$. Then the geodesics
which generate $S_\eps$ leave the point $P$, enter the region $D_\eps$
at $\kappa^\eps_{s,\gamma}(t^\eps_0)$ where
$t^\eps_0=-a/\sin\gamma+O(\eps)$, emerge from $D_\eps$ at 
$\kappa^\eps_{s,\gamma}(t^\eps_1)$ where
$t^\eps_1=-a/\sin\gamma+O(\eps)$, and then move outwards to
infinity. By the previous analysis the two portions outside $D_\eps$ 
both tend to straight lines as $\eps \to 0$, while the two points
$\kappa^\eps_{s,\gamma}(t^\eps_0)$ and $\kappa^\eps_{s,\gamma}(t^\eps_1)$ 
both tend to the point on the axis with Cartesian coordinates 
$(-a/\sin\gamma,0,0,-a\cot\gamma)$. Thus as $\eps \to 0$ such geodesics
tend to a straight line with a kink in it as it passes through the
axis. In Cartesian coordinates $\kappa^0_{s,\gamma}(t)$ is therefore
given by 
$$
\aligned
&\left.\aligned
x&=a+t\sin\gamma \\
y&=0  \\
z&=t\cos\gamma \endaligned\right\} \quad -a/\sin\gamma \leq t
\leq 0 \\\noalign{\smallskip}
&\left.\aligned
x&=-(a+t\sin\gamma)\cos\delta_s \\
y&=-(a+t\sin\gamma)\sin\delta_s  \\
z&=t\cos\gamma \\
\endaligned\right\} \quad t \leq -a/\sin\gamma
\endaligned
$$
where $\delta_s$ is some scattering angle which depends upon $s$.

For $s=1$ we have a geodesic which grazes $D_\eps$, so that its
deflection will be given by that of a geodesic in the conical
space-time, and hence $\delta_1=-\alpha$. Similarly $\delta_{-1}=\alpha$.
As $\gamma$ varies between zero and $\pi$ and $s$ varies between $-1$
and $1$, then the first part of the geodesic fills in the gap in 
$\hat C^-_0(P)$ caused by geodesics which hit the axis, while the
second part generates a portion of the surface
$$
x^2+y^2=(a-\sqrt{t^2-z^2})^2, \qquad t \leq -|z|
$$
with edges given by $y=\pm x\tan\alpha$ which exactly match with the tear
in $\hat C^-_0(P)$ and restore the $S^2$ topology of the null cone. See
Figure 2 for a section of the null cone through $t=\constant$,
$z=\constant$.

\midinsert
\medskip
\centerline{\epsfbox{pic2.ps }}
\medskip
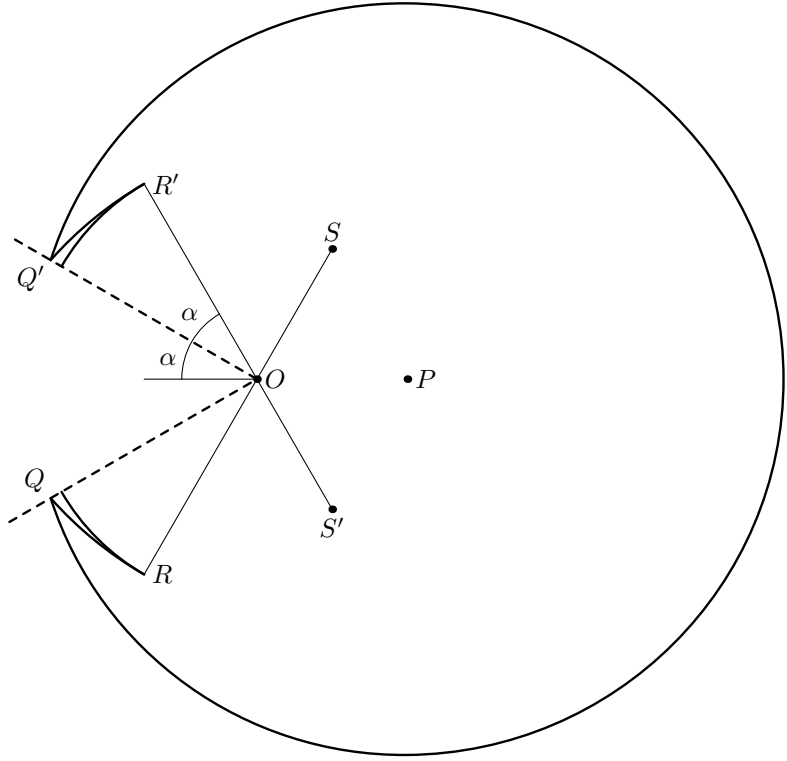
\centerline{\it Figure 2.}
\medskip
\endinsert

It is important to note that without imposing further conditions on
the regularisation (such as the scalar curvature of $g_\eps$ being
non-negative) we cannot assume that $\delta_s$ is a monotonic function
of $s$ or that it lies in the range $-\alpha \leq \delta_s \leq
\alpha$. However we do know that the boundary is given by
$\delta_{1}=-\alpha$ and $\delta_{-1}=\alpha$. Therefore if we
consider an integral of a scalar field over the parametrised surface
then (apart from a set of measure zero) each point outside the range
occurs an even number of times (with one half having the opposite
orientation from the other and so cancelling) and each point within 
the range occurs an odd number of times (with all but one occurrence
cancelling). So that if we integrate a scalar field over this region
we need only integrate over the region $\tilde C^-_0(P)$ given by
$$
x^2+y^2=(a-\sqrt{t^2-z^2})^2, \qquad t \leq -|z|, \quad
-x\tan\alpha \leq y \leq x\tan\alpha
$$
which is independent of the regularisation.

It is more natural to include the 2-surface generated by the first
part of the geodesic in the definition of $\hat C^-_0(P)$, so that we
can remove the restriction on geodesics hitting the axis  and $\hat C^-_0(P)$
is now simply defined to be the past null cone of $P$ in $(M \setminus
\Lambda)$. However~\eqref{intlim} remains valid, and combining this
with the above we obtain for all $\psi \in \cD(M)$
$$
\aligned
\lim_{\eps \to 0} \int_{C^-_\eps}(x)\psi(\xi)\mu^\eps_\Gamma(\xi) 
&=\int_{\hat C^-_0}(x)\psi(\xi)\hat \mu^0_\Gamma(\xi) 
+\int_{\tilde C^-_0}(x)\psi(\xi)\tilde \mu^0_\Gamma(\xi) \\
&=\int_{C^-_0}(x)\psi(\xi)\mu^0_\Gamma(\xi) \\
\endaligned
\tag\tagnum\xlabel{limint2}
$$
where $C^-_0(x)=\hat C^-_0(x) \cup \tilde C^-_0(x)$ and $\mu^0_\Gamma$
is the volume form induced by $g_0$ on $\hat C^-_0(x)$ and $\tilde C^-_0(x)$.
Note that points on the axis are excluded from both $\hat C^-_0(x)$ and 
$\tilde C^-_0(x)$, but do not contribute to the integral on the left
hand side either due to the (uniform) boundedness of $\mu^\eps$ even
when points on the axis are included.

The final point we have to deal with is to show that $\Lambda^-(x)$
and $\mu^0_{\Lambda^-}$ is well defined. $\Lambda^-(x)$ is defined to be
the limit of $C^-_\eps(x) \cap \Lambda$ as $\eps \to 0$. As we have
seen $\kappa^\eps_{s,\gamma}(t^0_\eps)$ tends to the point
$(-a/\sin\gamma,0,0,-a\cot\gamma)$ as $\eps \to 0$. So that as
$\gamma$ varies between
zero and $\pi$ we see that $C^-_\eps(P) \cap \Lambda$  is given by a
curve of the form
$$
x=0, \quad y=0, \quad t^2-z^2=a^2 +n_\eps(t,z), \quad t<0
\tag\tagnum\xlabel{limGamma}
$$
where $n_\eps \to 0$ uniformly as $\eps \to 0$. In the limit we obtain 
$\Lambda^-(x)$ which is given by
$$
x=0, \quad y=0, \quad t^2-z^2=a^2, \quad t<0
\tag\tagnum\xlabel{limGamma2}
$$
which is independent of the regularisation. 

Also for all values of
$\eps$ the volume form induced on $\Lambda$ by $g_\eps$ is given by
the 2-dimensional Minkowskian value $\mu_2$, (since the singular part is
orthogonal to the axis and the smoothing leaves the metric parallel to
the axis unchanged). Now by~\eqref{limGamma2} $\Lambda^-(x)$ is a smooth
curve so the volume form $\mu^0_{\Lambda^-}$ induced on it by $\mu_2$
is well defined and is indeed the limiting value of the volume form
induced on~\eqref{limGamma}.

\enddocument
\bye